\documentclass[%
 reprint,
superscriptaddress,
amsmath,amssymb,
 aps,
floatfix,
]{revtex4-2}

\usepackage{graphicx}% Include figure files
\usepackage{dcolumn}% Align table columns on decimal point
\usepackage{bm}% bold math
\usepackage{hyperref}% add hypertext capabilities
\usepackage{breqn}
\usepackage{subcaption}

\makeatletter
\let\cat@comma@active\@empty
\makeatother

\begin{document}

\title{Dynamic response of a ferromagnetic nanofilament under rotating fields: effects of flexibility, thermal fluctuations and hydrodynamics}% Force line breaks with \\

\author{Pedro A. S\'anchez}
\email{p.sanchez@uib.es}
\affiliation{Physics Department, University of the Balearic Islands, 07122 Palma, Spain.}

\author{Antonio Cerrato}
\affiliation{Physics Department, University of the Balearic Islands, 07122 Palma, Spain.}

\author{Joan J. Cerd\'a}
\affiliation{Physics Department, University of the Balearic Islands, 07122 Palma, Spain.}

\author{Carles Bona-Casas}
\affiliation{Physics Department, University of the Balearic Islands, 07122 Palma, Spain.}

\author{Tom\'as Sintes}
\affiliation{Instituto de F\'isica Interdisciplinar y Sistemas Complejos, IFISC (UIB-CSIC), University of the Balearic Islands, 07122 Palma, Spain.}

\author{Joan Mass\'o}
\affiliation{Physics Department, University of the Balearic Islands, 07122 Palma, Spain.}

%\date{\today}% It is always \today, today,
             %  but any date may be explicitly specified

\begin{abstract}
Using nonequilibrium computer simulations, we study the response of ferromagnetic nanofilaments, consisting of stabilized onedimensional chains of ferromagnetic nanoparticles, under external rotating magnetic fields. In difference with their analogous microscale and stiff counterparts, which have been actively studied in recent years, nonequilibrium properties of rather flexible nanoparticle filaments remain mostly unexplored. By progressively increasing the modeling details, we are able to evidence the qualitative impact of main interactions that can not be neglected at the nanoscale, showing that filament flexibility, thermal fluctuations and hydrodynamic interactions contribute independently to broaden the range of synchronous frequency response in this system. Furthermore, we also show the existence of a limited set of characteristic dynamic filament configurations and discuss in detail the asynchronous response, which at finite temperature becomes probabilistic.
\end{abstract}

%\keywords{Suggested keywords}%Use showkeys class option if keyword
                              %display desired
\maketitle

%\tableofcontents

\section{\label{sec:intro}Introduction}
Colloidal magnetic filaments (MFs) are microscopic structures created by stabilizing linear chains of magnetic micro- or nanoparticles with binding coatings, typically polymer crosslinkers. These chains combine a strong magnetic response with some degree of mechanical flexibility. \cite{2005-cebers-cocis} Due to their magnetoelastic properties, MFs can be used for a broad range of applications. For instance, they are one of the main approaches for the design of magnetically actuated microscopic swimmers and microrobots \cite{2005-dreyfus, 2009-belovs-pre, 2013-peyer, 2020-yang-pnas} and are used as micromechanical and microrheological sensors, actuators and mixers.\cite{2003-goubault-prl, 2003-wilhelm-pre1, 2003-wilhelm-pre2, 2009-fahrni-lc, 2013-chevry-pre, 2013-chen-lc, 2014-morozov}

Available experimental techniques make the synthesis of MFs readily from the crosslinking of magnetic microparticles with different properties. Most experimental works to date have been focused on paramagnetic or superparamagnetic microparticles,\cite{2004-biswal-pre, 2014-byrom, 2017-kuei-prf} typically formed by a soft material embedded with magnetic nanoparticles (MNPs), whereas microparticles consisting on a non-magnetic core coated with a layer of ferromagnetic material are used to create ferromagnetic filaments.\cite{2009-erglis-jmmm, 2010-erglis-mhd} The usage of microparticles allows a relatively high control of the crosslinking process and the tuning of the rigidity of the chains in a broad range, from very stiff to rather flexible.\cite{2003-biswal, 2010-li, 2014-byrom}

The downsizing of MFs to the nanoscale by stabilizing chains of individual MNPs can largely expand the range of applications of these systems.\cite{2011-wang} However, the direct crosslinking of MNPs is much more challenging, limiting most available experimental studies to non-crosslinked, magnetically self-assembled chains \cite{2010-benkoski-sm}, crosslinked semiflexible particle bundles \cite{2020-greiniankovski-jpcc, 2021-greinIankovski-em} or rather rigid, rod-like crosslinked chains.\cite{2007-benkoski-jacs, 2013-chevry-pre} The latter include not only synthetic materials but also polymer encapsulated chains of MNPs created by biomineralization within certain bacteria.\cite{1975-blakemore, 2008-faivre-chr} To this regard, cutting-edge techniques such as DNA directed assembly are paving the way to port the flexibility that can be achieved for crosslinked microparticles to nanofilament structures.\cite{2021-xiong-nl, 2022-mostarac-ns}

Following the development of the synthesis techniques, most existing theoretical studies on MFs are devoted to systems based on magnetic microparticles, with a main focus on the dynamic response and hydrodynamic properties. In most cases the filaments are assumed to be rather rigid and a continuum magnetoelastic description can be applied to systems with paramagnetic/superparamagnetic microparticles,\cite{2004-shcherbakov-pre, 2011-cebers-epje, 2017-vazquez-montejo-prm} ferromagnetic microparticles\cite{2008-erglis-jpcm, 2009-erglis-jmmm, 2010-erglis-mhd} or bacteria with embedded MNPs.\cite{2007-erglis-bpj} Such continuum descriptions incorporate magnetic torques and viscous frictions to the Kirchhoff model of nonstretchable elastic rods \cite{2003-cebers-jpcm, 2006-belovs-pre, 2011-javaitis-amr}. In addition, interparticle magnetic and hydrodynamic interactions and the effects of background thermal fluctuations are usually disregarded. Whereas all these approximations ease the fitting of the continuum models to achieve good agreement with experimental measurements, they make this approach inadequate to treat relatively high chain flexibilities and nanoscale structures.

The aforementioned limitations of magnetoelastic continuum descriptions can be overcome by means of bead-spring simulation models. These can incorporate naturally different approaches and levels of detail to describe the system, allowing to explore broad ranges of chain flexibilities and to consider the effects of thermal fluctuations and long-range hydrodynamic interactions with the only expected drawback of a larger computational cost. For instance, these models have been applied to both, microparticle chains\cite{2017-kuei-prf, 2022-spataforasalazar-jpcm} and MNPs nanofilaments, either rigid\cite{2023-camp-sms} or semiflexible. In the latter case, equilibrium properties have been largely addressed for ferromagnetic nanofilaments.\cite{2013-sanchez-jcp, 2015-sanchez-sm1, 2020-mostarac-ns, 2020-sanchez-pre} However, despite their great potential interest, studies on out of equilibrium properties of such latter systems are still very scarce due to the higher complexity and computing cost involved in its modeling, since long-range hydrodynamic interactions can not be disregarded under such conditions.\cite{2016-luesebrink-jcp, 2022-mostarac-mm} In this work we take a first step to fill such theoretical gap by studying how chain flexibility, background thermal fluctuations and hydrodynamic interactions affect the dynamics of ferromagnetic nanofilaments under an external drive. Importantly, none of these aspects can be generally disregarded for out of equilibrium conditions at the nanoscale.

For simplicity, here we consider one of the most elemental conditions among systems of MNPs under external drives: the application of a rotating magnetic field. Being broadly used to study the rheological properties of magnetic soft matter,\cite{2023-cebers} applied rotating fields have been the basis of numerous works on microparticle MFs.\cite{2004-cebers-pre, 2004-biswal-pre, 2017-goyeau-pre, 2017-kuei-prf, 2020-zaben-jmmm, 2020-zaben-sm} Moreover, they are one of the main ingredients in many of the practical applications mentioned above, thus the conditions we treat here are relevant for both, the fundamental characterization of the nanofilaments and their technological applications. With this goal, we set up a representative bead-spring model of a ferromagnetic nanofilament and study its dynamic response to rotating fields by means of extensive molecular dynamics simulations, showing the large impact of the aforementioned parameters on its behavior.

The paper is organized as follows. First, we introduce the reference physical system and the simulation model we use to represent it. Next, we present the simulation results for three increasing levels of modeling details: first, we analyze the athermal system in absence of hydrodynamic interactions, discussing the effects of field strength and filament rigidity; next, we focus on the limit of high filament flexibility under a moderate field strength and discuss the effects of finite temperature; finally, we also consider hydrodynamic interactions and discuss the impact of each interaction by direct comparison of the results with each set of model details. We conclude with a summary of the main findings.

\section{System, model and methods}

The system we consider is an ideal MF formed by a chain of equivalent nanospheres consisting of a monodomain ferromagnetic nanoparticle as a solid core and a polymer soft shell that stabilizes the particles and their chain-like arrangement. Chain stabilization may be achieved either by simple polymer crosslinking of the surfaces of neighboring particles\cite{2014-hill} or by means of pre-assembled polymer nanocages that encapsulate each MNP and remain connected by semiflexible joints.\cite{2021-xiong-nl, 2022-mostarac-ns} Such different stabilization techniques can lead to MFs with a broad range of bending rigidities, from very stiff in the first case to rather flexible in the second. We consider the MF to be immersed in a viscous Newtonian fluid and exposed to a rotating magnetic field of constant strength. The MNPs are assumed to have a size close to their monodomain limit, so that they possess a large magnetic anisotropy. Therefore, their magnetic moment can be represented as a point magnetic dipole fixed in their rigid body frame and oriented in a direction parallel to the chain backbone. The latter would correspond to perform the chain stabilization process under a constant applied field.

\begin{figure}[!h]
\centering
\includegraphics[width=0.85\columnwidth]{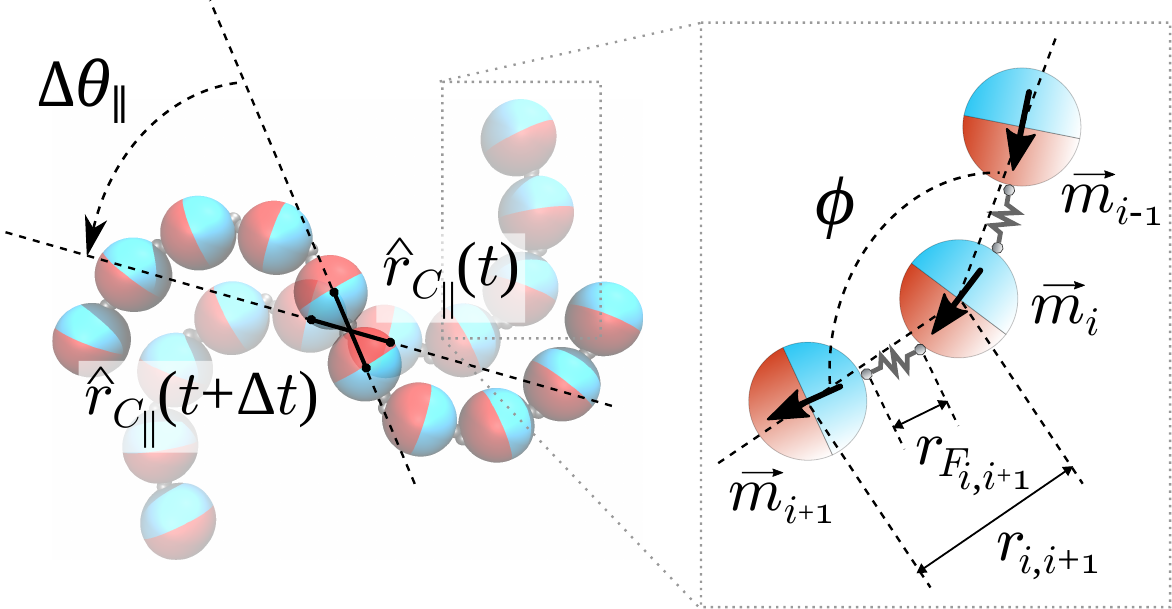}% Here is how to import EPS art
\caption{\label{fig:model}Main common parameters of the nanofilament model. MNPs are represented as spheres with different color hemispheres indicating the orientation of the magnetic dipole. Filament rotation velocity is measured from the component parallel to the applied field plane of the unit vector oriented along the axis connecting the centers of the two central particles.}
\end{figure}

In order to represent the system introduced above, we use a conventional bead-spring modeling approach,\cite{2020-sanchez-pre} whose main characteristics are summarized in the sketch shown in Figure~\ref{fig:model}. This approach is particularly convenient for the combined treatment of thermal fluctuations and interparticle magnetic and hydrodynamic interactions using molecular dynamics simulations. The effects of the thermal fluctuations of the carrier fluid can be represented implicitly by means of a stochastic thermostat. Since we consider our system to be formed by nanoparticles, we can approximate their dynamics by neglecting inertial effects and use a Brownian dynamics thermostat.\cite{2010-schlick} Whenever hydrodynamic interactions are not neglected, we combine the Brownian dynamics thermostat with a first order approximation for the hydrodynamic equations.\cite{2016-ran-njp} This defines for every particle $i$ in the system the following Brownian dynamics equations, which govern its linear and angular velocities, $\vec v _i$ and $\vec \omega_i$ respectively:
\begin{dmath}
 \vec v_i = \frac{\vec F_i^+}{6 \pi \eta R} + \\
 \sum_{j \ne i} \left ( \frac{1}{8 \pi \eta r_{ij}} \left [ \vec F_j^+ + \frac{1}{r_{ij}^2} \left ( \vec F_j^+ \cdot \vec r_{ij} \right ) \vec r_{ij} \right ] - \frac{\vec r_{ij} \times \vec T_j^+}{8 \pi \eta r_{ij}^3}\right ), \label{eq:BrownT}
 \end{dmath}
 \begin{dmath}
 \vec \omega_i = \frac{\vec T_i^+}{8 \pi \eta R^3} + \\
 \sum_{j \ne i} \left ( \frac{1}{8 \pi \eta} \left [ - \frac{\vec T_j^+}{r_{ij}^3} + \frac{3}{r_{ij}^5} \left ( \vec T_j^+ \cdot \vec r_{ij} \right ) \vec r_{ij} \right ] - \frac{\vec r_{ij} \times \vec F_j^+}{8 \pi \eta r_{ij}^3}\right ) \label{eq:BrownR}.
 \end{dmath}
Here, $r_{ij} = \| \vec r_{ij} \| = \| \vec r_i - \vec r_j \|$ is the length of the center-to-center displacement vector between particles $i$ and $j$, $\eta$ the dynamic viscosity of the background fluid, $R$ the hydrodynamic radius of the particles and $\vec F _i^+$ and $\vec T_i^+$ are the net force and torque acting on $i$, including thermal fluctuations:
\begin{eqnarray}
 \vec F ^+_i = {\vec F}_i + (12 \pi \eta R\, kT)^{1/2}\,{\hat \xi}_{i,\mathrm{T}},\label{eq:netF}\\
 {\vec T}^+_i = {\vec T}_i + (16 \pi \eta R^3\, kT)^{1/2}\,{\hat \xi}_{i,\mathrm{R}},\label{eq:netT}
\end{eqnarray}
being $\vec F_i$ and $\vec T_i$ the net conservative force and torque, respectively. The remaining terms represent the thermal fluctuations, which are introduced as a stochastic force and torque with random orientation, given by the unit vectors with random components ${\hat \xi}_{i,\mathrm{T}}$ and ${\hat \xi}_{i,\mathrm{R}}$, and a variance defined by $\eta$, $R$, and the thermal energy of the system, given by the product of the Boltzmann constant, $k$, and the absolute temperature, $T$.

Note that the first terms in the right hand side of Equations~\ref{eq:BrownT} and \ref{eq:BrownR} correspond to the classical Brownian dynamics terminal linear and angular velocities,\cite{2010-schlick} with the denominators being the Stokes translational and rotational friction coefficients for a sphere in a viscous fluid. The remainig terms represent a rough, first order approximation to hydrodynmaic interactions in the form of the Oseen tensor with rotational components. These dynamic equations have been used recently to study small systems of MPNs under rotating magnetic fields in the limit of vanishing thermal fluctuations\cite{2016-ran-njp, 2020-ran-itm}.

The point dipole representation allows to treat explicitly interparticle magnetic interactions using the conventional dipole-dipole long-range pair potential:
\begin{equation}
U_{\mathrm{dip}}(\vec m_i;\, \vec m_j;\, \vec r_{ij})= \\ \frac{\mu_0}{4 \pi} \left [ \frac{\vec{m}_i\cdot\vec{m}_j}{r_{ij}^{3}}- \right .
\left  . \frac{3\left[\vec{m}_i\cdot\vec{r}_{ij}\right]\left[\vec{m}_j\cdot\vec{r}_{ij}\right]}{r_{ij}^{5}} \right ],
\label{eq:dipdip}
\end{equation}
where $\vec{m}_{i}$ and $\vec{m}_{j}$ are the dipole moments of particles $i$ and $j$ and $\mu_0$ is the magnetic permittivity of vacuum, thus we assume a non-magnetic background with relative permittivity $\sim$1. Additionally, the dipole moment of each particle experiences the Zeeman interaction with the applied external field, $\vec H$, according to
\begin{equation}
 U_{\mathrm{Z}}(\vec m_i,\, \vec H) =- \vec m_i \cdot \mu_0 \vec H.
 \label{eq:Zeeman}
\end{equation}
The field is applied in the $x$--$y$ plane, with constant strength $H$ and rotation frequency $f_H$:
\begin{equation}
 \vec H(t) = H \left [ \cos \left ( 2 \pi f_H t \right ),\, \sin \left ( 2 \pi f_H t \right ),\,  0\right ].
 \label{eq:field}
\end{equation}

Finally, we choose a minimal implicit representation of the excluded volume and chain stabilization effects, keeping in mind that the molecular dynamics approach requires smooth potentials for all conservative interactions. Thus, we represent the nanoparticles forming the filament as soft spheres with excluded volume interactions defined by a truncated and shifted Lennard-Jones potential, also known as Weeks-Chandler-Andersen (WCA) soft-core potential \cite{1971-weeks-jcp}:
\begin{equation}
U_{\mathrm{{WCA}}}(r_{ij})=\left\{ \begin{array}{ll}
U_{\mathrm{{LJ}}}(r_{ij}) - U_{\mathrm{{LJ}}}(r_{\mathrm{cut}}), & r_{ij}<r_{\mathrm{{cut}}}\\
0, & r_{ij}\geq r_{\mathrm{{cut}}}
\end{array}\right. ,
\label{eq:WCA}
\end{equation}
where $U_{\mathrm{LJ}}(r_{ij}) = 4\epsilon_{\mathrm{s}} \left[\left(2R/r_{ij}\right)^{12} - \left(2R/r_{ij}\right)^{6}\right]$ is the conventional Lennard-Jones potential and $r_{\mathrm{cut}} = 2^{7/6}R$. The chain connectivity is ensured by a bonding potential that links points on the surface of neighbor particles corresponding to the projection of the axes defined by their dipole moments at a distance $R$ from their centers (see Figure~\ref{fig:model}). The chosen bonding interaction is the finitely extensible non-linear elastic (FENE) bonding potential\cite{1972-warner-iecf}:
\begin{equation}
 U_{\mathrm{FENE}}(r_F) = -\frac{1}{2} K_F r_{\mathrm{max}}^2  \ln \left [ 1 - \left ( \frac{r_F}{r_{\mathrm{max}}} \right ) ^2\right ] ,
 \label{eq:FENE}
\end{equation}
where $K_F$ is the elastic strength of the bond, $r_F$ is its length and $r_{\mathrm{max}}$ its maximum extension. Note that this potential, in combination with \ref{eq:WCA}, mainly limits the separation between neighbor particles along the chain to a given range. Even though to apply the bonding potential on the surface of the particles tends to increase the penalty of chain bending with respect to that of a simple chain of self-avoiding spheres free to perform rolling motions around each other, this effect would be large only for a combination of very steep soft core potential and very small maximum bond length, whereas in any other case the chain will remain rather flexible. Thus, in order to easily tune the bending rigidity of the filament in a broad range, we introduce an additional explicit bending potential. For simplicity, we take a simple harmonic form:
\begin{equation}
 U_{b} (\phi) = \frac{K_b}{2} \left ( \phi - \pi \right )^2,
 \label{eq:bending}
\end{equation}
where $K_b$ is the strength of the bending penalty and $\phi$ is the angle defined by three consecutive neighbor particles along the chain (see Figure~\ref{fig:model}).

In order to set the model parameters, we choose a representative reference physical system. This is a filament formed by $N=10$ magnetite (Fe$_2$O$_4$) nanoparticles, with a saturation magnetization of $M_s = 400$~kA/m, hydrodynamic radius of $R=20$~nm and a polymer layer of negligible thickness. This corresponds to a particle magnetic moment of $\| \vec m\| \approx 1.34 \cdot 10^{-17}\ \mathrm{A} \cdot \mathrm{m}^2$. This filament is immersed in water at room temperature and exposed to rotating fields in broad ranges of strengths and rotation frequencies, respectively, $H \in [50,\, 400]$~kA/m and $f_H \in [20,\, 400]$~kHz. In molecular dynamics simulations, however, it is convenient to use a system of dimensionless units that rescales the physical parameters around handling values. This serves to both, ensure the numerical stability of the integration of the dynamic trajectories and to ease the transferability of the results to other systems with the same ratios of their physical parameters. For instance, the results presented below could also correspond to an analogous filament made of cobalt nanoparticles of $R=13$~nm with a non-magnetic coating of 7~nm thickness.  The parameter scales chosen to discuss the results presented below are the following. Energies will be measured in terms of the thermal energy at room temperature, $kT_{\mathrm{room}}$. The magnetic field strength scale will be given by the saturation magnetization of the magnetic material forming the particles, $M_s$. As length scale we take the contour length of the filament, $L$, which corresponds to its end-to-end distance for an ideal, perfectly straight configuration:
\begin{equation}
L = (N - 1) b,
\label{eq:lengthsc}
\end{equation}
where $b$ is the center-to-center distance between first nearest neighbors along the chain. Finally, as time scale we will take a factor relating the characteristic strength of the viscous and elastic forces in the system, equivalent to the prefactor of the so-called elasto-viscous number,\cite{2018-liu-pnas} frequently used in continuum MF models:\cite{2017-goyeau-pre,2023-stikuts-jmmm}
\begin{equation}
t_s = 8 \pi \eta L^4 / B_0,
\label{eq:timesc}
\end{equation}
where $B_0$ is an arbitrary reference value of the filament bending rigidity, $B$. Note that, in general, the bending rigidity of a polymer-like chain is related to its persistence length, $l_p$, by the well known relationship\cite{1986-doi}
\begin{equation}
 B = l_p\, kT.
 \label{eq:Blp}
\end{equation}
For $B_0$ we take the lowest value that can be sampled in our model, \textit{i.e.}, the rigidity of our model filament in absence of the bending potential \ref{eq:bending}.

At this point we formally introduced all the interactions and the choices for the system scales used in our simulations. The physical parameters introduced above fully define expressions \ref{eq:dipdip}, \ref{eq:Zeeman} and \ref{eq:field}, as these correspond to explicit models of physical interactions. However, the implicit representation of the chain stabilization given by expressions \ref{eq:WCA}, \ref{eq:FENE} and \ref{eq:bending} still requires a mapping to the physical properties of the filament, represented by the parameters $b$ and $B_0$ in expressions \ref{eq:lengthsc} and \ref{eq:timesc}. In other words, neither $b$ nor $B$ can be set directly in our model, but depend in a non-trivial way on all the interactions introduced above. However, we can take advantage of the fact that $b$ and $l_p$ can be directly estimated from simulation data: by means of separated preparatory simulations at zero field and room temperature (not shown), we mapped a set of model parameters to filament physical properties in a range of interest. First, the combination \ref{eq:WCA} and \ref{eq:FENE} sets a bond in which the distance between the linked points and, thus, the center-to-center neighbor distance, $b \equiv r_{i, i+1}$, can fluctuate around an average value $\langle b \rangle$. The choice of parameters for \ref{eq:WCA} and \ref{eq:FENE} can be arbitrary as long as they lead to small fluctuations around $\langle b \rangle \approx 2R$ while not imposing a very small integration time step. We found the set $\epsilon_s / kT_{\mathrm{room}}= 100$, $K_F / kT_{\mathrm{room}} = 3000$ and $r_{\mathrm{max}} = R$ to fulfill these requirements. By taking the measured $\langle b \rangle$ as estimator of $b$ in expression \ref{eq:lengthsc}, we established the reference length scale as $L \approx 18.4R$. Second, we mapped the strength of the bending potential \ref{eq:bending} to the effective bending rigidity of the filament by sampling a wide range of values $K_b/kT_{\mathrm{room}} \in \left [ 0,\, 10^4 \right ]$ and measuring the corresponding effective persistence length. The latter can be done by fitting the bond angle correlation function of the filament configurations, $g_{\mathrm{bac}}$,\cite{1986-doi} to a decaying exponential,
\begin{equation}
 g_{\mathrm{bac}} (n) = \left \langle \frac{\vec b_{i,1} \cdot \vec b_{i , n} }{\left \| \vec b_{i,1}\right \| \left \| \vec b_{i,n}\right \|} \right \rangle_i \propto \exp \left ( - \frac{2nR}{l_p} \right ),
 \label{eq:l_p}
\end{equation}
where $\vec b_{i,n} = \vec r_{i + n} - \vec r_i $ is the center-to-center displacement vector between $n$-th nearest neighbors along the chain. As expected,\cite{1986-landau-lifshitz} we found a linear relationship between $l_p$ and $K_b/kT_{\mathrm{room}}$:
\begin{equation}
 l_p \approx \frac{2K_b R}{kT_{\mathrm{room}}} + l_p^0.
\end{equation}
For the maximum sampled value $K_b/kT_{\mathrm{room}} = 10^4$ we obtained $l_p / L \approx 1100$, whereas for $K_b/kT_{\mathrm{room}} = 0$ we mesaured $l_p^0/L \approx 10$. This non-zero value of $l_p^0$ has two contributions: as pointed above, the finite stretchability of the bonds connecting the surfaces of neighbor particles provides some rigidity to the backbone; in addition, dipole-dipole interactions between particles along the chain also contribute to increase it.\cite{2013-sanchez-jcp} The relative importance of these contributions can be roughly estimated from the persistence length of a chain of free dipolar particles,\cite{2004-morozov} which in a first order approximation is estimated as $l_p^d \approx \mu_0 \| \vec m \| ^2 / (32 \pi kT R^2)$. In our case $l_p^d/L \approx 4$, thus we can conclude that the bonds and the dipole-dipole interactions contribute to a similar extent to the filament rigidity. For the physical reference system, the range of effective persistence lengths we obtained is $l_p \in [3.5 \cdot 10^{-6},\, 4\cdot 10^{-4}]$ m. According to expression \ref{eq:Blp}, this corresponds approximately to effective bending rigidities from $B \approx 10^{-26}\ \mathrm{kg \cdot m^3 / s^2} \equiv B_0$ to $B \approx 10^{-24}\ \mathrm{kg \cdot m^3 / s^2}$. Note that this upper limit is close to the lowest experimental rigidities reported for paramagnetic microparticle filaments.\cite{2004-biswal-pre,2014-byrom} Morevoer, typical bending rigidities studied in most works based on continuum models are at least four orders of magnitude larger than the highest value sampled here.\cite{2004-cebers-pre, 2008-erglis-jpcm, 2009-erglis-jmmm} Thus, with our approach we can achieve the goal of expanding those studies to the limit of small filament rigidities. Finally, with the effective value we determined for $B_0$, the time scale for the physical reference system is set to $t_s \approx 1.2 \cdot 10^{-3}\ \mathrm{s}$.

The simulation protocol used to obtain the results presented below is the following. The filament was initially placed with an ideally straight configuration parallel to plane $x$--$y$ and with the same initial orientation as the applied external field. The dynamic equations were integrated with a symplectic velocity Verlet algorithm using a time step of the same order of the Brownian relaxation time of the particles forming the filament, which for the reference physical system corresponds to $\delta t \approx 5\cdot 10^{-10}$~s. The rotation of the field was simulated by discretizing its orientations according to the selected rotation frequency and the integration time step, ensuring that several integration steps were performed for each discrete field orientation in order to prevent discretization artifacts and, for simulations at finite temperature, to achieve a proper system thermalization. The simulations were run for $10^3$ field turns, in the case of athermal simulations, and for $5 \cdot 10^3$ for thermalized systems. In the latter case, statistical distributions were obtained from up to 50 independent runs for each set of parameters. All the simulations were performed using the simulation package ESPResSo 4.2.1,\cite{2019-weik-epjst} with a custom extension implementing the first order approximation of hydrodynamic interactions given in expressions \ref{eq:BrownT} and \ref{eq:BrownR}.

\section{Results and discussion}
In order to unveil the effects of rigidity, thermal fluctuations and hydrodynamic interactions on the response of our model filament to rotating fields, we performed separate sets of simulations with increasing levels of model details, adding each of such parameters one by one. Thus, in the next sections we first discuss the ideal athermal case with a simple background viscous friction to determine the role of the filament rigidity below the rigid rod limit; next, we introduce thermal fluctuations to analyze their effects and, finally, we add hydrodynamic interactions to conclude with an overall comparison of the response under these different conditions.

%SMALL BUT FINITE STRETCHABILITY OF THE BONDS -> SMALL DEVIATIONS OF THE DIPOLE MOMENT FROM THE EASY AXIS

\subsection{Effects of filament rigidity: athermal system}
In order to analyze in detail the effects of the filament rigidity we start with a simplified model in which thermal fluctuations and hydrodynamic interactions are disregarded. Note that, by disregarding long-range hydrodynamic interactions, equations \ref{eq:BrownT} and \ref{eq:BrownR} are reduced to their first term only, whereas the athermal approximation makes expressions \ref{eq:netF} and \ref{eq:netT} to not include the stochastic terms. Under these conditions, trajectories are constrained to the plane parallel to the field and strictly follow deterministic dynamics, thus we will label the results of this simple model as `DD'.

\subsubsection{Frequency response}

\begin{figure}[h]
\centering
\includegraphics[width=0.95\columnwidth]{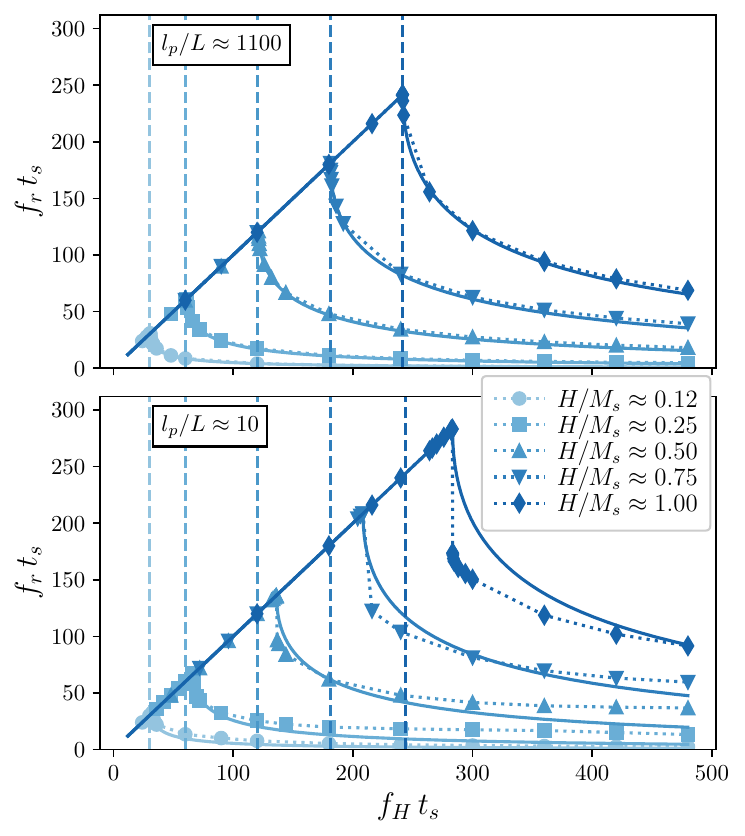}% Here is how to import EPS art
\caption{Frequency response of DD model filaments, $f_r t_s$, as a function of the field frequency, $f_H t_s$, for the two extreme sampled persistence lengths, $l_p/L \approx 1100$ and $l_p/L \approx 10$, and different field strengths, $H/M_s$. Solid lines correspond to the ideal response given by Eq.~\ref{eq:syncasync}, fitted to the maximum of each simulation data set. Vertical dashed lines indicate the corresponding critical frequency of an ideal rigid filament, $l_p/L \rightarrow \infty$,  given by Eq.~\ref{eq:criticalrod} with $R=\langle b \rangle/2$. Symbols correspond to simulation data, with uncertainty intervals of the order of symbol size, dotted lines connecting symbols are a guide for the eye.\label{fig:fresp-DD}}
\end{figure}

In general, any small body with a net magnetic moment that is immersed in a viscous fluid and exposed to a rotating magnetic field\cite{2006-mcnaughton-prb, 2017-goyeau-pre} displays two main dynamic regimes: in the \emph{synchronous regime} the rotation frequency of the body, $f_r$, is the same of that of the field, $f_H$, up to a given critical frequency, $f_H \le f_c$; for bigger field frequencies, $f_H > f_c$, an \emph{asynchronous regime} takes place due to the dominance of viscous friction, which leads to a strong drop of the effective rotation frequency of the body with respect to $f_H$. For rigid bodies constrained to rotate in the plane of the field, their frequency response is given by\cite{2006-mcnaughton-prb}
\begin{equation}
 f_r(f_H) = \left\{ \begin{array}{ll}
f_H, & f_H \le f_c\\
f_H - \left ( f_H^2 - f_c^2 \right )^{1/2}, & f_H > f_c
\end{array}\right. .
\label{eq:syncasync}
\end{equation}
Ideally, the critical frequency is given by the balance between the net magnetic and the net viscous torques, $\tau_m$ and $\tau_\gamma$ respectively, which act oppositely on the body. In therms of the ratio
\begin{equation}
 M_a \equiv \tau_\gamma / \tau_m,
 \label{eq:mason}
\end{equation}
which is known as the magnetorheological Mason number,\cite{2003-melle-pre} the critical frequency corresponds to the condition $M_a \approx 1$.\cite{2009-gauger-epje} In the limit of infinite rigidity, the critical frequency of our model filament can be simply estimated as
\begin{equation}
 f_c^\infty \equiv {\left ( f_c \right )}_{l_p \rightarrow \infty} = \frac{ \left \| \vec m \right \| \mu_0 H}{4 \eta (N^2 - 1) \pi^2 R^3}.
 \label{eq:criticalrod}
\end{equation}
However, we can expect increasing deviations from the ideal rigid rod structure as the filament flexibility grows. Note that expressions \ref{eq:syncasync} and \ref{eq:criticalrod} can be made dimensionless by simply multiplying the frequencies by the time scale, $t_s$.

Figure~\ref{fig:fresp-DD} shows the dimensionless frequency response, $f_r\, t_s$, of DD model filaments corresponding to both extreme sampled values of the room temperature persistence length and several field strengths, together with the ideal response curves given by Equation~\ref{eq:syncasync}, fitted with the maximum measured value of $f_r\, t_s$ as critical frequency, $f_c\, t_s \equiv \mathrm{max}(f_r\, t_s)$, and the critical frequencies of the infinitely rigid filament given by \ref{eq:criticalrod} when taking the measured average center-to-center neighbor distance as an estimator of the effective particle radius, $R \approx \langle b \rangle/ 2$. As observed for continuum models of rather rigid microfilaments,\cite{2017-goyeau-pre} our results show an excellent agreement with the theoretical expressions \ref{eq:syncasync} and \ref{eq:criticalrod} for the most rigid case. However, in the response curves of the most flexible case two significant deviations can be observed: a shift towards higher values of the critical frequencies that grows with increasing field strength and a qualitatively different behavior in the asynchronous region, with a more abrupt drop of the response immediately above the critical value and a smaller asymptotic decay.

\begin{figure}[!t]
\centering
\includegraphics[width=0.95\columnwidth]{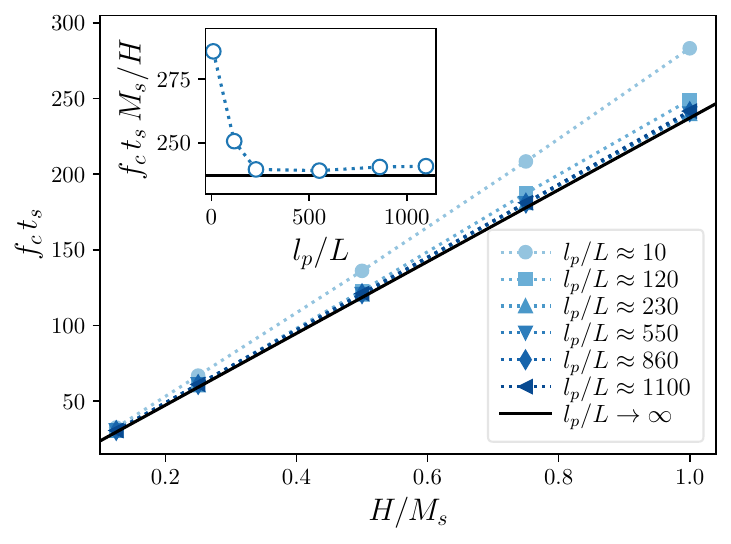}% Here is how to import EPS art
\caption{\label{fig:fc-BD-T0}Dependence of the critical frequency of the DD model, $f_c\,t_s$, on field strength, $H/M_s$, for different filament rigidities, $l_p/L$ (main figure) and field-rescaled critical frequency, $f_c\,t_s\,M_s/H$, as a function of the filament rigidity (inset). Solid lines correspond to the ideal rigid rod-like case, $l_p/L \rightarrow \infty$.}
\end{figure}
Figure~\ref{fig:fc-BD-T0} provides a closer look at the effect of rigidity on the critical frequencies, showing their dependence on the field strength for the whole range of sampled rigidities. As reference, it also includes the values given by \ref{eq:criticalrod} for the fully rigid filament. The main plot evidences that in all cases $f_c\, t_s$ grows linearly with the field, with a slope that tends to increase with the filament flexibility. The behavior for large rigidities tends to that of the fully rigid case. This can be better observed in the inset, which shows the value of the slope $f_c \, t_s\, M_s/H$ as a function of $l_p/L$. This dependence is strongly nonlinear, with a large drop of the slope as the rigidity grows up to $l_p / L \lesssim 230$, whereas for higher rigidities it remains basically constant and very close to the value of the fully rigid case. This independence from rigidity in its limit of large values is consistent with the observation of a constant proportionality between the critical frequency and the so-called magnetoelastic number, $C_m \sim H/l_p \propto f_c$, found in continuum models for rather rigid microfilaments.\cite{2017-goyeau-pre} Thus, with this we identified the region for which the effects of the chain flexibility become important and the rigid filament approximation is not valid anymore.

\begin{figure*}[!h]
\centering
\begin{subfigure}[b]{0.95\columnwidth}
\centering
\includegraphics[width=0.95\columnwidth]{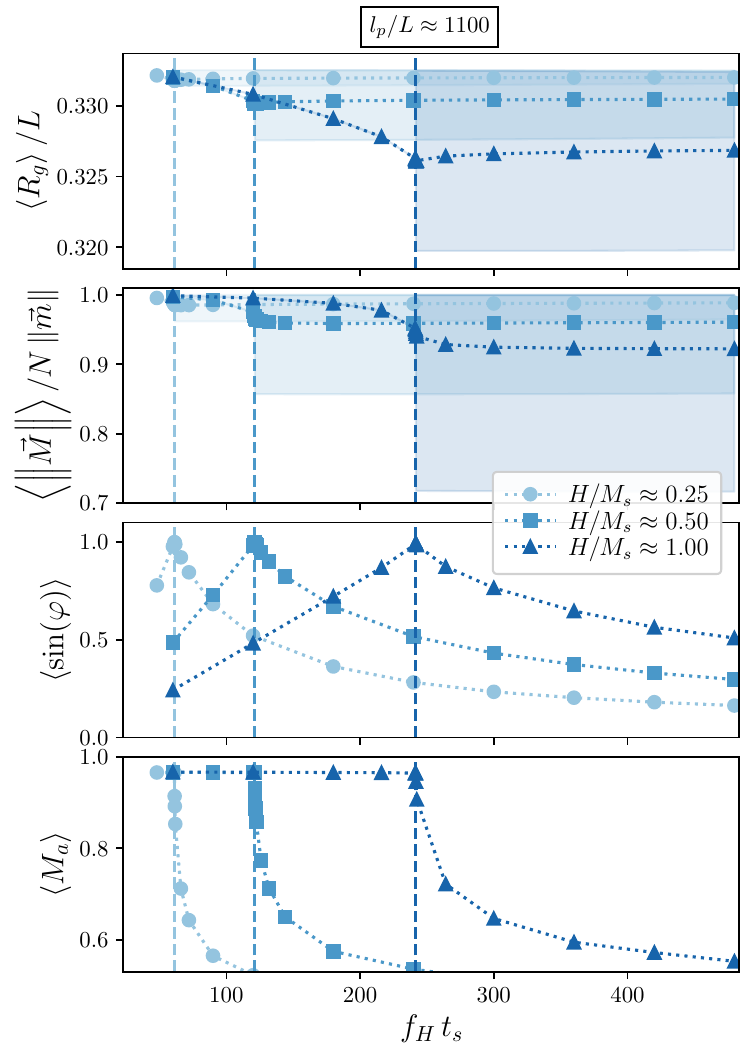}% Here is how to import EPS art
\caption{\label{fig:structresp-BD-Kb10000-T0}}
\end{subfigure}
\begin{subfigure}[b]{0.95\columnwidth}
\centering
\includegraphics[width=0.95\columnwidth]{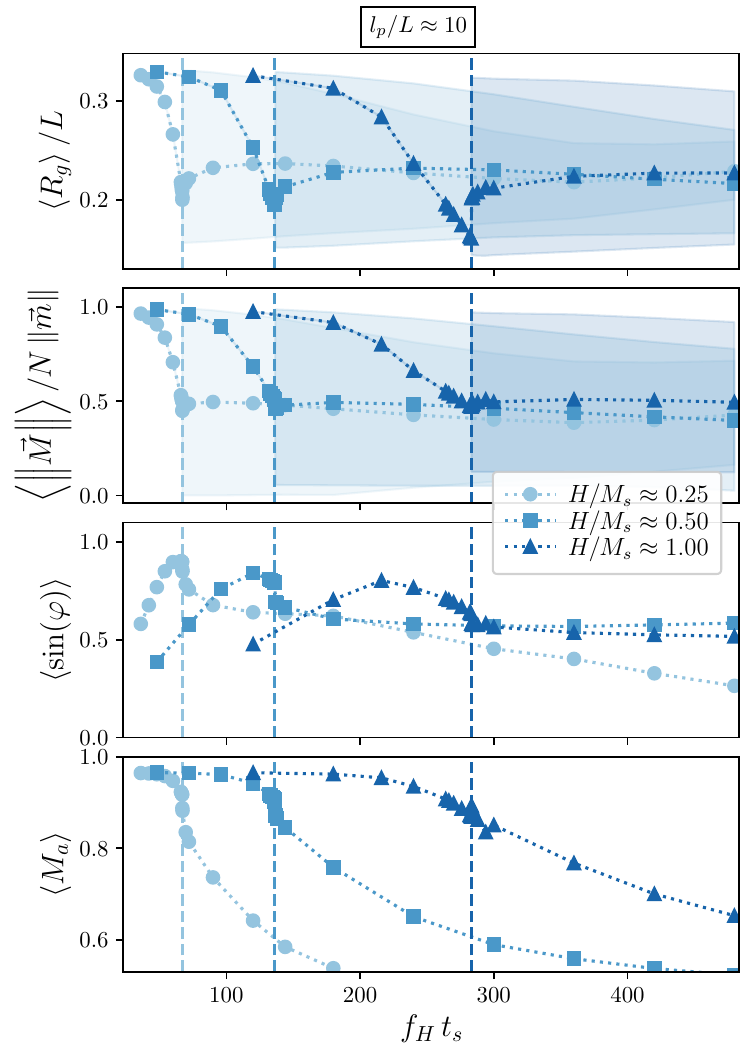}% Here is how to import EPS art
\caption{\label{fig:structresp-BD-Kb0-T0}}
\end{subfigure}
\caption{\label{fig:structresp-BD-T0}Structural and magnetoviscous parameters as a function of the field frequency, $f_H\,t_s$, for three selected field strengths, $H/M_s$, and the highest (a) and lowest (b) sampled values of filament rigidity, $l_p/L$: average radius of gyration, $\left \langle R_g \right \rangle / L$, average net magnetization, $\left \langle \left \| \vec M \right \| \right \rangle / N \left \| \vec m \right \| $, average phase shift, $\left \langle \sin\left ( \varphi \right) \right \rangle $, and average Mason number, $\left \langle M_a \right \rangle$. Vertical dashed lines correspond to the critical frequencies. Note the different vertical scales used in (a) and (b) for radius of gyration and net magnetization. Shadowed regions for the latter parameters indicate the range of values explored due to structural oscillations in the asynchronous regime.}
\end{figure*}
The deviations of the frequency response observed in the limit of high filament flexibility can be reasonably attributed to characteristic deformations of the filament structure with respect to the ideal straight configuration. In order to analyze the interplay between filament structure and its response to the rotating field we compute an accurate estimator of the dimensionless Mason number \ref{eq:mason} from the simulation data. First, we calculate the viscous torque by assuming that the filament rotates at any time around its center of mass, located at $\vec r_{\mathrm{CM}}$. For a given effective rotation frequency $f_r$--- which can be calculated from the simulation trajectories by linear regression of the time accumulated in-plane rotation angle, $\Delta \theta_{\parallel}(t)$, see Figure~\ref{fig:model}---each particle forming the filament contributes to the net viscous torque depending on its distance to $\vec r_{\mathrm{CM}}$:
\begin{equation}
 \tau_{\gamma} \approx 12 \pi^2 \eta R f_r \sum_{i=1}^N (\vec r_i - \vec r_{\mathrm{CM}})^2 = 12 \pi^2 \eta R f_r N R_g^2,
 \label{eq:frictorque}
\end{equation}
where $R_g$ is the radius of gyration of the filament. Second, the net magnetic torque is given by the cross product of the net magnetic moment of the filament, $\vec M = \sum_{i=1}^N \vec m_i$, and the applied field:
\begin{equation}
 \tau_m = \mu_0 \left \| \vec M \times \vec H \right \|= \mu_0 \left \| \vec M \right \| \left \| \vec H \right \| \left | \sin \varphi \right |,
 \label{eq:magtorque}
\end{equation}
where $\varphi$ is the phase shift between the net magnetic moment and the field. Thus,
\begin{equation}
 M_a = \frac{12 \pi^2 \eta R f_r N R_g^2}{\mu_0 \left \| \vec M \right \| \left \| \vec H \right \| \left | \sin \varphi \right |}.
 \label{eq:masondetail}
\end{equation}
Figure~\ref{fig:structresp-BD-T0} shows the evolution of the time averages of this estimation of the Mason number and its parameters as a function of the dimensionless field frequency, for the two extreme sampled filament rigidities and three selected field strengths. Importantly, the separation between the synchronous and asynchronous regimes is clearly signaled in such evolution for any rigidity: specifically, the onset of the asynchronous regime is associated to the appearance of significant time fluctuations of the parameters, illustrated for the relative radius of gyration, $R_g/L$, and relative net magnetization, $\| \vec M \| / (N \left \| \vec m \right \|)$. Before analyzing such fluctuations, we first focus on the differences that arise depending on the rigidity.

Similarly to what is expected for a rigid rod, for the most rigid sampled filament (left panel of Figure~\ref{fig:structresp-BD-T0}) the Mason number remains very close to 1 for the synchronous region. This is associated to a large growth of the phase shift and a slight decay of the radius of gyration and net magnetization up to the critical frequency. At the latter, the phase shift reaches its maximum, $\varphi \approx \pi/2$. This means that in the synchronous regime the increase of the viscous friction torque with the growth of the angular velocity is mainly compensated with an increase of the phase shift, which enhances the magnetic torque, whereas the rigidity of the filament only allows small deformations of the straight configuration. For the asynchronous regime, the structural parameters indicate that the filament shape slightly fluctuates around a configuration close to that at the critical frequency, whereas the average phase shift and Mason number tend to drop as the enhancement of the asynchrony broadens the range of phase shifts explored by the filament in each turn of the field.

Despite showing qualitative features similar to the ones of the most rigid case, the behavior of the most flexible filament (right panel of Figure~\ref{fig:structresp-BD-T0}) is noticeably more complex. First, a much larger drop of the radius of gyration and net magnetization in the synchronous region indicate important changes in the shape of the filament. Second, the structural fluctuations in the asynchronous region do not average to the configuration at the critical frequency. Finally, the maximum of the phase shift and the starting point of the drop of the Mason number are observed at frequencies slightly below the critical one. Thus, in order to understand these differences we need a more detailed look at the evolution of the filament structure in each case. In the next Sections we present such details, splitting the discussion for the synchronous and asynchronous regimes for clarity.

\subsubsection{Synchronous regime: stationary structures}
\begin{figure}[!t]
\centering
\begin{subfigure}[b]{0.95\columnwidth}
\centering
\includegraphics[width=0.7\columnwidth]{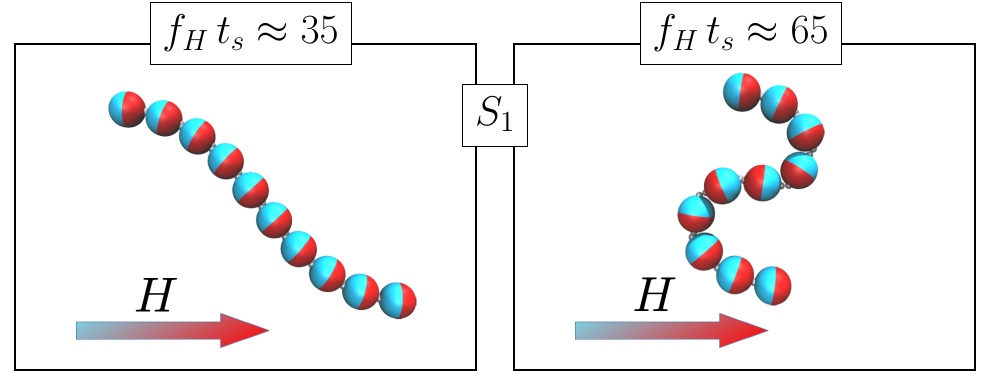}
\caption{\label{fig:S1-snaps}Snapshots showing the compaction of the $S_1$ configuration with field frequency for the most flexible sampled filament, $l_p/L \approx 10$.}
\end{subfigure}
\begin{subfigure}[b]{0.95\columnwidth}
\centering
\includegraphics[width=0.95\columnwidth]{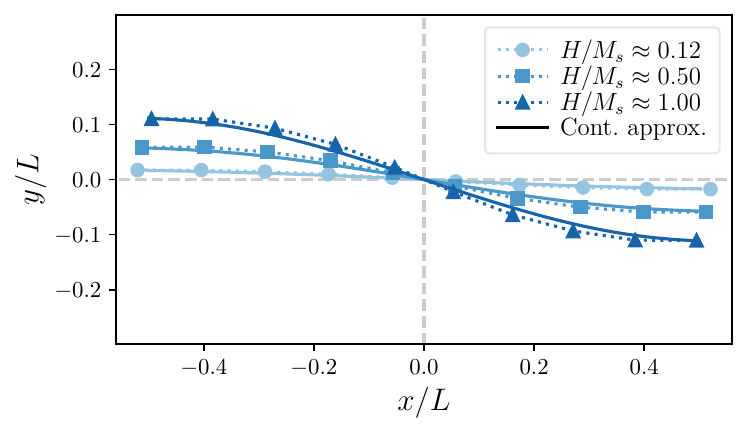}% Here is how to import EPS art
\caption{\label{fig:shapes-fc-Hs-BD-T0}Comparison of the filament backbone shape obtained for the most rigid sampled filament, $l_p/L \approx 1100$, at the critical frequency and different field strengths (symbols) to the predictions given by Equation~\ref{eq:analshapes}, derived from a continuum description (solid lines). Symbols correspond to the positions of the center of the particles, with the filament center of mass placed at the origin and the free ends parallel to the reference axis, $x$.}
\end{subfigure}
\caption{Examples of the effect of field frequency (a) and strength (b) on the $S_1$ shapes of the DD model under synchronous conditions.}
\label{fig:S1shapes}
\end{figure}

The results presented in Figure~\ref{fig:structresp-BD-T0} indicate that for a fixed set of parameters---filament rigidity, field strength and field frequency---the filament adopts a stationary shape under synchronous regimes. In all cases, this is a simple ``S'' shape with different degrees of curvature, as illustrated by the profiles shown in Figure~\ref{fig:S1shapes}. This characteristic morphology is found in continuum models of very rigid ferromagnetic microparticle filaments,\cite{2017-goyeau-pre} as well as experimentally in their paramagnetic counterparts.\cite{2017-kuei-prf} Hereinafter, we will refer to this shape as ``$S_1$''.

Figure~\ref{fig:S1-snaps} shows two snapshots of the $S_1$ shapes adopted by the most flexible filament in the synchronous regime at different field frequencies. In general, the decrease of the radius of gyration with increasing field frequency discussed in Figure~\ref{fig:structresp-BD-T0} corresponds to an increase of the local curvature and compaction of the $S_1$ morphology. Besides this, the increase of the field strength shifts the critical frequency towards higher values, allowing to reach more compact $S_1$ shapes in the synchronous regime. The latter is illustrated in Figure~\ref{fig:shapes-fc-Hs-BD-T0}, which presents several characteristic stationary shapes adopted at critical frequencies by the most rigid filament under different field strengths. These profiles are also compared to the predictions of a model for the $S_1$ structures that was deduced very recently from the continuum representation of rather rigid microfilaments.\cite{2023-stikuts-jmmm} In such approach, small deviations of the filament profile from a reference straight rod configuration, from which one can define a parallel reference axis $x$, are given by the parametric equation
\begin{dmath}
  y(l) = L \frac{\Delta y}{\Delta x} \left [ \frac{6 \sinh \left ( l \sqrt{C_m} / L \right )}{C_m \sinh \left ( \sqrt{C_m} / 2 \right )} - 2 \left ( \frac{l}{L}\right )^3 + \frac{l}{L} \left ( \frac{3}{2} - \frac{12}{C_m}\right )\right ].
 \label{eq:analshapes}
\end{dmath}
where $y$ is the distance to the reference axis, $l \in [-L/2,\, L/2]$ is the archlength parameter along the filament, $C_m$ is the aforementioned magnetoelastic dimensionless number and $\Delta y / \Delta x$ is the ratio of the filament contour lengths measured, respectively, in perpendicular and parallel directions to the reference axis while assuming that the free ends remain parallel to the latter. In order to compare our results to this model, we set $\Delta y/\Delta x$ directly from the shapes obtained in our simulations. Regarding $C_m$, the approximations assumed by the continuum descriptions, on which expression \ref{eq:analshapes} is based, make difficult to obtain an analogous meaningful definition for $C_m$ within our model. However, for the sake of comparison we can take advantage of the fact that for any $C_m \gg 1$ Equation~\ref{eq:analshapes} becomes rather insensitive to this parameter. With the aforementioned fitting of the prefactor $\Delta y / \Delta x$, we found that any $C_m > 10^3 H/M_s$ provides negligible differences in the predicted profiles, being this limit the best approximation to the shapes obtained in our simulations. The examples shown in Figure~\ref{fig:shapes-fc-Hs-BD-T0} illustrate the general good agreement of our results with the large $C_m$ limit of Equation~\ref{eq:analshapes}, but they also evidence that deviations become more significant as the field strength and, thus, the deformation with respect to the ideal straight configuration, are increased.
\begin{figure}[!h]
\centering
\begin{subfigure}[b]{0.95\columnwidth}
\centering
\includegraphics[width=0.95\columnwidth]{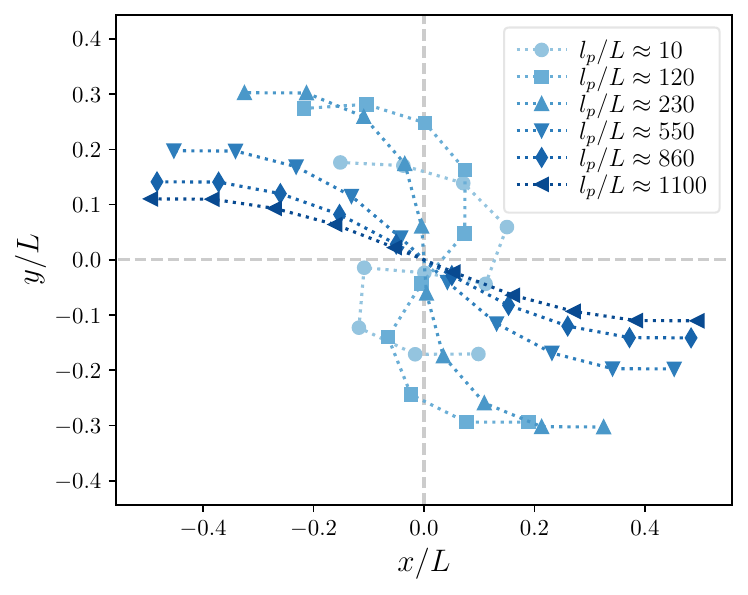}
\caption{\label{fig:shapes-fc-Kbs-BD-T0}Influence of the filament rigidity on its shape at the critical frequency and field strength $H/M_s \approx 1$. Same representation as in Figure~\ref{fig:shapes-fc-Hs-BD-T0}.}
\end{subfigure}
\begin{subfigure}[b]{0.95\columnwidth}
\centering
\includegraphics[width=0.95\columnwidth]{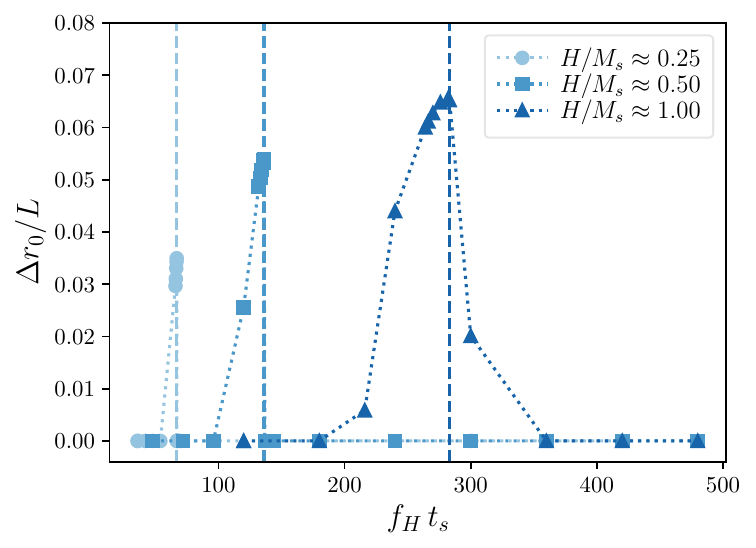}
\caption{\label{fig:COMshifts-fH-BD-T0}Relative shift of the center of rotation (center of mass) with respect to the chain geometrical center (middle point between particles 5th and 6th), $\Delta r_{\mathrm{O}} / L$, as a function of the field frequency for $l_p/L \approx 10$ and different field strengths.}
\end{subfigure}
\caption{Evidence of slight asymmetries in the $S_1$ shape around the critical frequency in the limit of high flexibility.}
\label{fig:BD-T0-flex-shapes}
\end{figure}

As the flexibility of the filaments increases, their deformations under the synchronous regime become too large to follow Equation~\ref{eq:analshapes}. Figure~\ref{fig:BD-T0-flex-shapes} evidences the large impact of this parameter on the synchronous behavior. The profiles of filaments with different rigidities, under strong field at critical frequencies, are shown in Figure~\ref{fig:shapes-fc-Kbs-BD-T0}. Here, one can see not only the large compactions experienced by the most flexible systems, which clearly explain the large drops of the radius of gyration in Figure~\ref{fig:structresp-BD-Kb0-T0} as the critical frequency is approached, but also the appearance of an unexpected asymmetry in the profiles. This asymmetry shifts the center of mass of the configurations, $\vec r_{\mathrm{CM}}$, away from the geometrical center of the ideal straight shape, which corresponds to the middle point between the two central particles along the chain, $\vec r_{\mathrm{O}} = (\vec r_5 + \vec r_6)/2$. This shift, denoted as $\Delta r_{\mathrm{O}} = \| \vec r_{\mathrm{CM}} - \vec r_{\mathrm{O}} \|$, is quantified in Figure~\ref{fig:COMshifts-fH-BD-T0} for the same system parameters discussed in Figure~\ref{fig:structresp-BD-Kb0-T0}, \textit{i.e.}, the most flexible case under three field strengths and the whole sampled range of field frequencies. In all cases the shift appears as the field frequency approaches its critical value, reaching a maximum at $f_c$ and dropping very fast to zero with the onset of the asynchronous regime. Interestingly, the height and width of these peaks grow with the field strength. Moreover, this peak broadening happens within the same regions in which large drops in the phase shift and Mason number right below the critical frequencies are observed in Figure~\ref{fig:structresp-BD-Kb0-T0}. Therefore, the latter can be explained by the onset of shape asymmetries. Whereas the technical reason for these asymmetries to appear in deterministic simulations can be simply the limited numerical accuracy inherent to any computer numerics, the fact that they become the stationary configurations in the system can be attributed to the higher ratio of the profile compaction, which decreases the overall viscous friction, to the associated decrease of net magnetization they provide. In this way, asymmetries shift the critical frequencies towards higher values than the ones allowed by strictly symmetric configurations. Importantly, these asymmetries do not simply represent a numeric artifact, as they could be triggered in experimental systems by filament inhomogeneities or external pertubations, thus they may represent a true physical effect. Deviations from ideal conditions in the properties or the initial setup of microparticle filaments have been suggested as the origin of several of their dynamic structural properties.\cite{2011-erglis-jmmm,2020-zaben-sm} However, to our best knowledge, the particular behavior observed here was not reported in studies existing to date, as it emerges from a range of flexibilities that was not yet explored.

\subsubsection{Asynchronous regime: structural oscillations}
Shape oscillations of relatively stiff microparticle filaments under asynchronous regimes have been observed in experiments, continuum models and athermal computer simulations.\cite{2004-cebers-pre,2017-kuei-prf,2023-spataforasalazar-sm} However, to the best of our knowledge, this is the first time that are reported for ferromagnetic nanoparticle filaments. Thus, here we check whether the change in scale and the exploration of a wider range of flexibilities brings in significant differences.
\begin{figure}[!b]
\centering
\begin{subfigure}[b]{0.95\columnwidth}
\centering
\includegraphics[width=0.95\columnwidth]{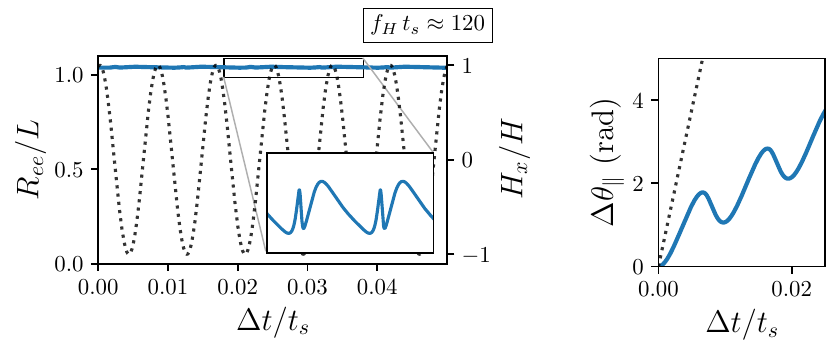}% Here is how to import EPS art
\caption{\label{fig:asyncosc-BD-T0-Kb10000}Time evolution of the end-to-end distance and accumulated rotation angle for the most rigid sampled filament, $l_p/L \approx 1100$.}
\end{subfigure}
\begin{subfigure}[b]{0.95\columnwidth}
\centering
\includegraphics[width=0.95\columnwidth]{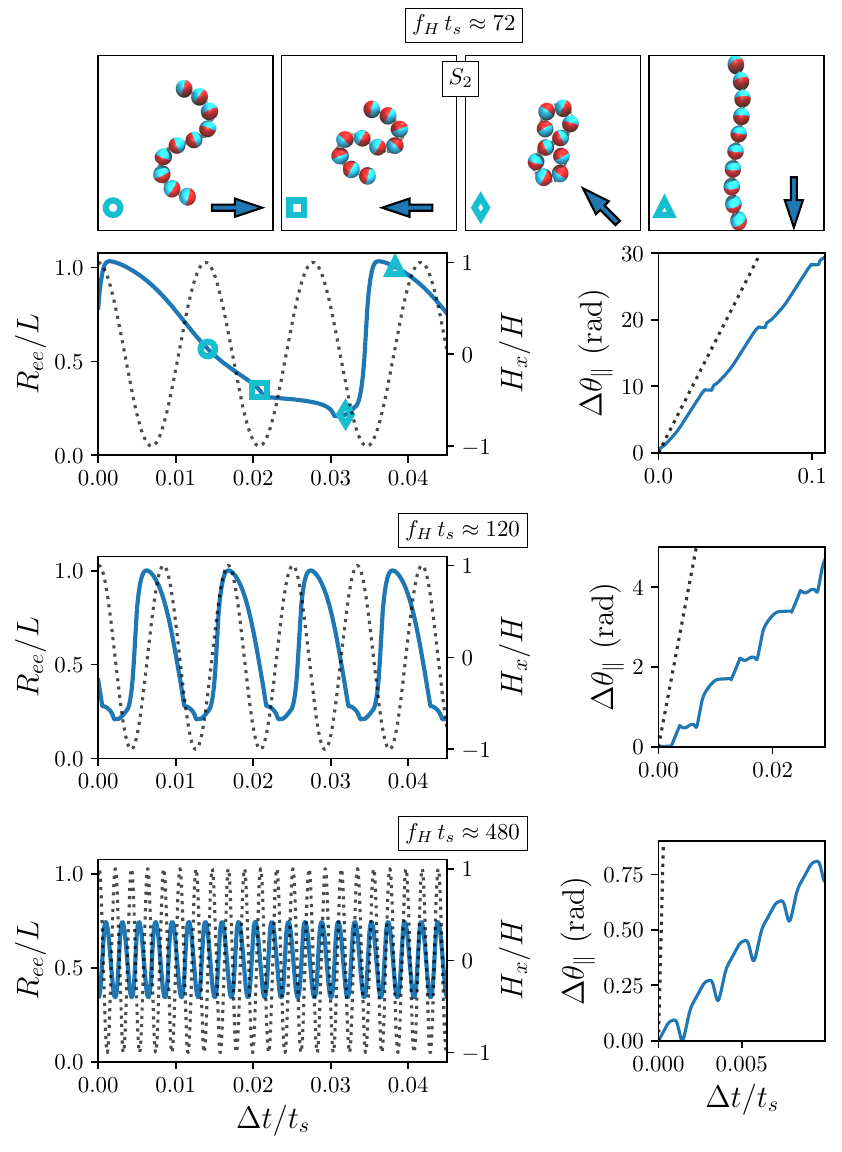}% Here is how to import EPS art
\caption{\label{fig:asyncosc-BD-T0-Kb00000}Time evolution of the end-to-end distance and accumulated rotation angle for the most flexible sampled filament, $l_p/L \approx 10$, for three selected field frequencies. Configuration snapshots at different points of the structural oscillations are included for the lowest frequency.}
\end{subfigure}
\caption{\label{fig:asyncosc-BD-T0}Details of the structural oscillations of $S_2$ configurations in the DD model under $H/M_s \approx 0.25$ and asynchronous conditions, for the limits of high (a) and low (b) filament rigidities: time evolution of the filament end-to-end distance, $R_{ee}/L$, and accumulated rotation angle, $\Delta \theta_{\parallel}$, (solid lines) compared, respectively, to the component of the field along a reference axis $x$, $H_x/H$, and the field accumulated rotation angle (dotted lines).}
\end{figure}

Figure~\ref{fig:asyncosc-BD-T0} shows in detail the nature of the oscillations we observe for $H/M_s \approx 0.25$. These are characterized by means of the time evolution of the filament end-to-end distance, $R_{ee}$, together with the time accumulated rotation angle, $\Delta \theta_\parallel$, and several snapshots. As already indicated by the results presented in Figure~\ref{fig:structresp-BD-T0}, these fluctuations depend strongly on the filament rigidity. Figure~\ref{fig:asyncosc-BD-T0-Kb10000} evidences that shape oscillations are very small in the limit of high rigidities. Under these conditions the filament keeps the characteristic $S_1$ shape while performing the typical asymmetric back and forth partial turns associated to driven asynchronous rotation regimes.\cite{2005-cebers-cocis,2007-erglis-bpj,2013-chevry-pre,2017-goyeau-pre} However, in the limit of small rigidities, illustrated by Figure~\ref{fig:asyncosc-BD-T0-Kb00000}, shape oscillations at field frequencies  moderately above the critical value can be so large that the structure periodically changes between a rather extended and a very compact closed double loop ``S'' shapes, with an oscillation period significantly lower than the one of the field. This behavior resembles an extreme case of the `wagging' oscillations reported in continuum models\cite{2004-cebers-pre}, athermal MD simulations and experiments\cite{2017-kuei-prf,2023-spataforasalazar-sm} of paramagnetic microparticle filaments, for the case in which the limit of `coiling' of the free ends is reached. As the field frequency grows, the dynamics becomes too fast to allow large expansions and compactions of the filament structure, thus the range of the oscillations tends to shrink around moderately compact ``S'' shapes and their period tends to approach the one of the field, whereas the net rotation rate of the filament keeps decreasing. Finally, by increasing the field strength, the general behavior described above tends to shift towards higher field frequencies. In the following we will refer to this oscillating structure as ``$S_2$''.

%SIMILARITY COMES FROM LIMITED EQUIVALENCE LENGTH-FLEXIBILITY (TO CHECK THEIR RIGIDITY), BUT VERY LONG STRUCTURES ARE DIFFICULT TO SYNTHESIZE AT THE NANOSCALE

\subsection{Effects of thermal fluctuations}
The first step towards a more realistic representation of nanoparticle filaments is to add the thermal forces \ref{eq:netF} and \ref{eq:netT} to the dynamic equations of the system, \ref{eq:BrownT} and \ref{eq:BrownR}, performing simple Brownian dynamics simulations, \textit{i.e.}, without considering hydrodynamic interactions. Hereafter we will label this simple thermalized model as `BD'. Since the qualitative effects of rigidity and field strength have been already discussed for the DD model, hereafter we focus on the most flexible case, $l_p/L \approx 10$, under a moderate field strength, $H/M_s \approx 0.25$, sampling field frequencies only. Under these conditions, we expect the impact of thermal fluctuations on the filament response to be rather important.

\subsubsection{Long time configurations}
\begin{figure}[!h]
\centering
\includegraphics[width=0.95\columnwidth]{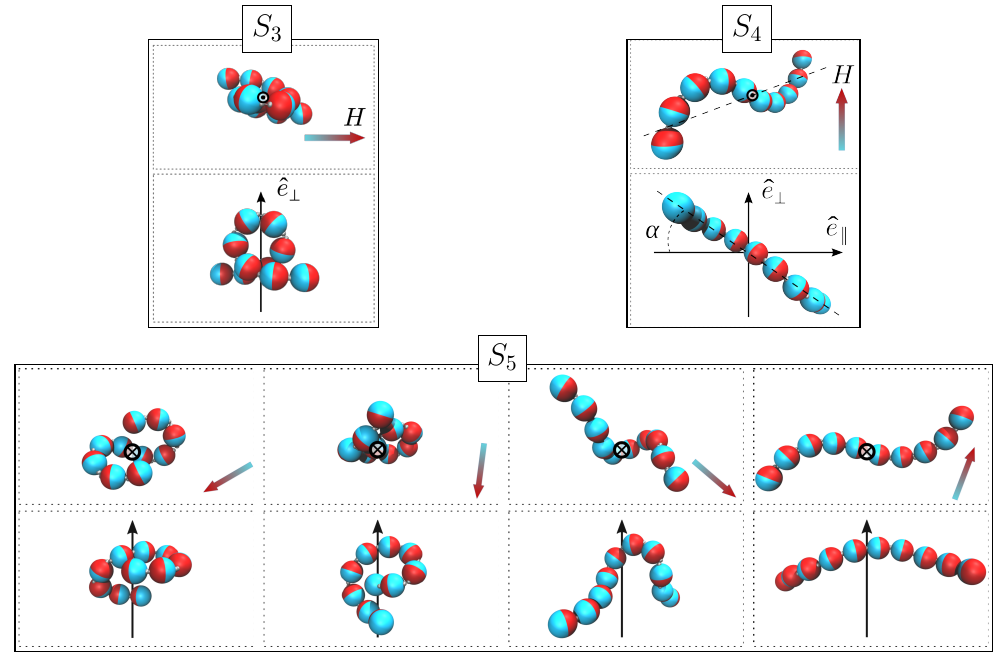}
\caption{\label{fig:snaps-BD-T1}Snapshots illustrating the long-time stationary structures, $S_3$, $S_4$ and $S_5$, observed at finite temperature in BD and HD models. For $S_5$, snapshots at different times are shown to evidence the large structural oscillations of this configuration.}
\end{figure}

In presence of thermal fluctuations it is reasonable to expect twofold main new effects: first, the thermalization may perturb and/or enhance the structural oscillations observed in the athermal system, producing statistical variations of the structural relaxation to different stationary configurations; second, thermal fluctuations provide force components perpendicular to the field rotation plane that may lead to the adoption of out of plane configurations. The latter assumption is confirmed by a simple inspection of the simulation trajectories: for all sampled field frequencies, after a very short transient the filament escapes out of the field rotation plane. Specifically, three new types of persistent configurations, displayed in Figure~\ref{fig:snaps-BD-T1}, appear as the shapes to which the thermalized filament relaxes at long times.
All these configurations have an out of plane component, while keeping the free ends parallel to the rotation plane.

The first type of out of plane configuration, which we label as ``$S_3$'', is a quasi-planar shape in which the central region of the filament forms an out of plane loop (see upper left panel in Figure~\ref{fig:snaps-BD-T1}). Similar shapes have been reported experimentally for semiflexible filaments of paramagnetic microparticles as a large elastic buckling induced by rotating fields.\cite{2004-biswal-pre} For ferromagnetic microparticle filaments, however, to date $S_3$ structures have been reported only as transient shapes under uniaxial field inversion.\cite{2010-erglis-mhd} The second out of plane configuration is a planar ``S'' shape that, in difference with $S_1$, forms a nonzero angle with the field rotation plane (shown in the upper right panel of Figure~\ref{fig:snaps-BD-T1}). This structure, that we label as ``$S_4$'', has been observed experimentally for asynchronous regimes of ferromagnetic microparticle filaments under rotating fields.\cite{2020-zaben-sm} Finally, there is an additional configuration we observe, ``$S_5$'', similar to $S_3$ but with a wider central loop. The latter enables short segments at the filament free ends to perform full rotations following the field, whereas the central part rotates much more slowly (see lower panel of Figure~\ref{fig:snaps-BD-T1} for a sequence of snapshots illustrating such rotations). To our best knowledge, this configuration has never been reported before. This, together with the large dynamic variation of shape it involves, suggests that it is possible only for large chain flexibilities.

\subsubsection{Frequency response of the thermalized system}
\begin{figure}[!h]
\centering
\begin{subfigure}[b]{0.95\columnwidth}
\centering
\includegraphics[width=0.95\columnwidth]{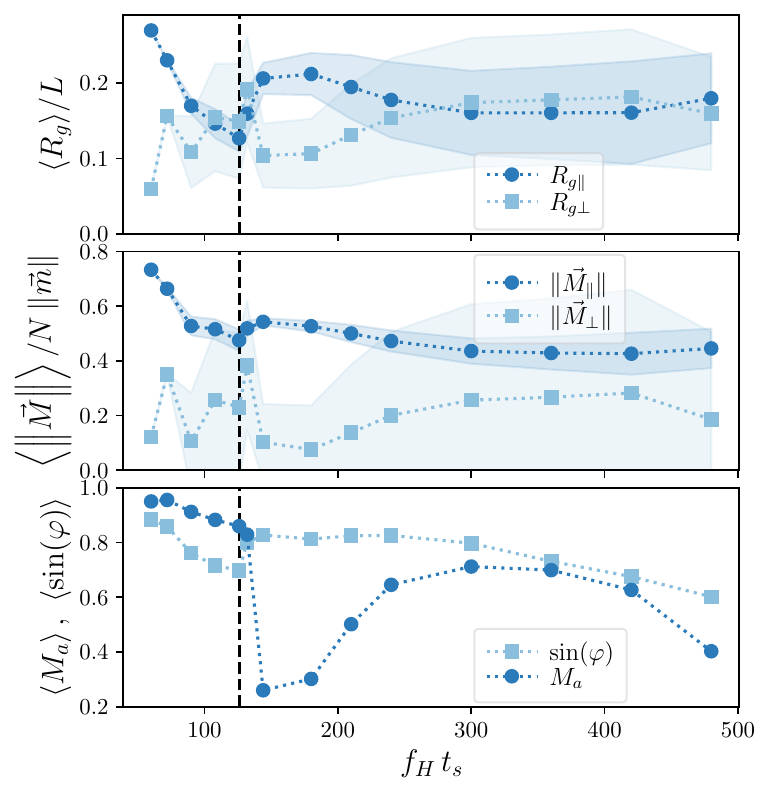}
\caption{\label{fig:structresp-BD-Kb0-T1-statparams}Long-time dependence of the structural and magnetoviscous parameters on field frequency. Symbols correspond to average values, shadowed regions to their standard deviation due to statistcal variations of the selected configurations. Radius of gyration and net magnetization are displayed with split parallel and perpendicular components with respect to the field rotation plane. Vertical dahsed lines indicate the critical frequency.}
\end{subfigure}
\begin{subfigure}[b]{0.95\columnwidth}
\centering
\includegraphics[width=0.99\columnwidth]{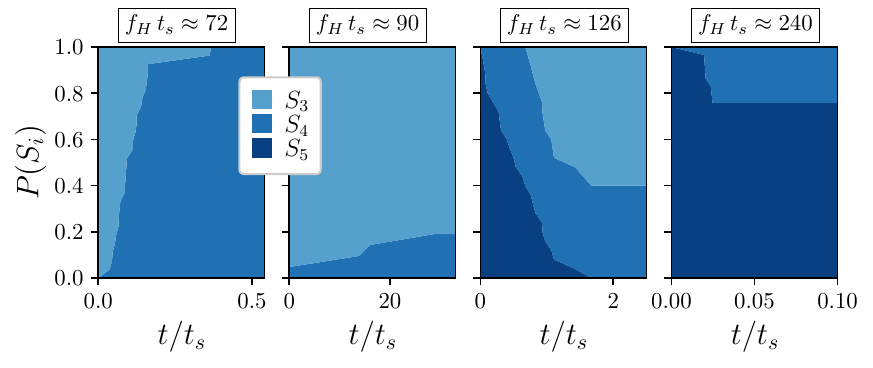}
\caption{\label{fig:structresp-BD-Kb0-T1-relaxdyn}Time evolution of the probability fractions of configurations $S_3$, $S_4$ and $S_5$, illustrating the statistics of the relaxation dynamics towards final long-time configurations for four selected field frequencies. $f_H\,t_s \approx 126$ corresponds to the critical frequency. In each case, only a limited early timeframe in which significant changes are observed is shown. }
\end{subfigure}
\caption{\label{fig:structresp-BD-Kb0-T1}Long-time frequency structural response of the BD model for a moderate field strength, $H/M_s \approx 0.25$, and low rigidity $l_p/L \approx 10$.}
\end{figure}
The overall frequency response of the BD filament model is presented in Figure~\ref{fig:structresp-BD-Kb0-T1}. In order to better characterize the out of plane configurations, the main structural parameters---radius of gyration and net magnetization---are shown in Figure~\ref{fig:structresp-BD-Kb0-T1-statparams} with split parallel and perpendicular components with respect to the field rotation plane, $x$--$y$. For the radius of gyration, such split components are calculated from the gyration tensor of filament configurations, $S$, whose elements are given by
\begin{equation}
 S_{(\alpha, \beta)} = \frac{1}{N} \sum_{i=1}^N \left ( r_{(\alpha),i} - \langle r_{(\alpha)} \rangle \right ) \left ( r_{(\beta),i} - \langle r_{(\beta)} \rangle \right ).
\end{equation}
With $\alpha,\, \beta \equiv \lbrace x,\, y,\, z \rbrace$, the eigenvalues of $S$, $\lbrace \lambda_1,\, \lambda_2, \, \lambda_3\rbrace$, give the spherically averaged radius of gyration as
\begin{equation}
 R_{g} = \left ( \lambda_1 + \lambda_2 + \lambda_3 \right )^{1/2}.
\end{equation}
By taking the coordinates in the field rotation plane only, $\alpha,\, \beta \equiv \lbrace x,\, y \rbrace$, the corresponding eigenvalues of $S$, $\lbrace \lambda_1,\, \lambda_2 \rbrace$ give the parallel component of the radius of gyration,
\begin{equation}
 R_{g\parallel} = \left ( \lambda_x + \lambda_y \right )^{1/2},
\end{equation}
whereas the perpendicular component is then obtained as
\begin{equation}
 R_{g \perp} = (R_g^2 - R_{g\parallel}^2)^{1/2}.
\end{equation}
For the net magnetization we simply calculate  \begin{equation}
 \left \| \vec M_{\parallel} \right \| = \left [ \left ( \sum_{i=1}^N m_{x,i} \right )^2 + \left ( \sum_{i=1}^N m_{y,i} \right )^2 \right ]^{1/2}
\end{equation}
and
\begin{equation}
 \left \| \vec M_{\perp}\right \| = \sum_{i=1}^N m_{z,i}.
\end{equation}
The first important observation concerns the critical frequency, which we define in the most general way as the highest value of the field frequency that strictly leads to synchronous rotations with all probability. The value measured here, $f_c \, t_s \approx 126$, is around 1.9 times bigger than the one of the corresponding athermal system (see Figure~\ref{fig:structresp-BD-Kb0-T0}). Thus, the adoption of out of plane configurations promoted by thermal fluctuations extends significantly the range of the synchronous regime. The critical frequency is signaled, respectively, by absolute and local minima of the average parallel components of the radius of gyration and net magnetization. In the synchronous regime, such parameters show a significant decrease of their averages with field frequency, with rather small fluctuations. However, the average Mason number and phase shift decrease only slightly, remaining close to unity. This behavior is qualitatively similar to the one observed for the athermal system except for the phase shift, which seems strognly enhanced by the thermal fluctuations. The behavior of the perpendicular components of the structural parameters, however, is way more intriguing, with large jumps in the averages and broad fluctuations that grow with frequency. This is due to the fact that, in difference to the DD system, here the fluctuations of the averages do not only correspond to periodic oscillations of a given structure, but also to the adoption of rather different shapes in independent simulation runs. Thus, at long times the system relaxes to one of the configurations $S_3$ to $S_5$, following frequency dependent statistical probabilities. Interestingly, we observed that at long times only the $S_3$ and $S_4$ configurations are adopted in the synchronous regime, whereas for any field frequency significantly above the critical limit only $S_4$ and $S_5$ are found. For the latter case, we can observe a large drop of the average out of plane components of radius of gyration, magnetization and Mason number right above the critical frequency, followed by smooth changes of these parameters at higher frequencies. Besides this, a more detailed discussion of the asynchronous regime is provided in the next Section, in the context of a comparative overview of the configurational probability distributions for each filament model. Here we mainly focus on the strongly irregular frequency dependence observed in the synchronous regime.

Below the critical frequency, the long time fractions of $S_3$ and $S_4$ configurations change nonmonotonically with $f_H$. At low frequencies the filament relaxes to $S_4$ only, increasing with $f_H$ its angle with respect to the rotation plane. This corresponds to the initial growth of $\langle R_{g\perp} \rangle$ in Figure~\ref{fig:structresp-BD-Kb0-T1-statparams}. However, at intermediate frequencies $S_3$ becomes dominant. The fact that this configuration has a smaller out of plane component than $S_4$ leads to the large drop of $\langle R_{g\perp} \rangle$, together with a large increase of its fluctuations, around $f_H\, t_s \approx 90$. Finally, as the frequency approaches its critical value, the fractions of $S_3$ and $S_4$ configurations become very similar, making $\langle R_{g\perp} \rangle$ to grow, approaching its maximum value. In order to understand these changes in the ratio of $S_3$ and $S_4$ long time configurations, we have to take a look at the relaxation dynamics leading to them.

In all of our simulations, the filaments change very quickly their initial straight configuration to one from the established set $S_1$ to $S_5$, typically at times $t/t_s < 10^{-3}$. However, we found that an additional single transition to a different configuration may happen at later times. Figure~\ref{fig:structresp-BD-Kb0-T1-relaxdyn} shows the time evolution of the fractions of $S_3$, $S_4$ and $S_5$ obtained from independent simulation runs for the BD model under selected low, intermediate and high values of the field frequency within the synchronous regime. At low frequency the filament initially adopts the $S_3$ configuration, but transitions to $S_4$ within a relatively short time span, $t/t_s < 0.5$, remaining as that for the maximum sampled time, $t/t_s \approx 40$. This suggests that $S_4$ is the most stable configuration under the combined action of thermal fluctuations and slow varying torques. At intermediate frequency, a small fraction of samples initially adopt and remain with the $S_4$ configuration, whereas a large fraction still adopt $S_3$ initially. Among the latter, only a small part transition to $S_4$ within a rather large time interval. This indicates that more fastly varying torques tend to difficult the transition $S_3$ to $S_4$. Finally, at high frequency, which in this case corresponds to the critical value, $f_c\, t_s \approx 126$, the filament is unable to relax initially to $S_3$ or $S_4$, adopting $S_5$ instead. Note that this configuration performs asynchronous rotations of the central part of the filament, which is the reference we use to distinguish the synchronous/asynchronous regimes. However, all initial $S_5$ configurations end up transitioning to either $S_3$ or $S_4$ in a similar proportion within a moderate time interval. Interestingly, no transition from $S_4$ to $S_3$ has been observed at any frequency. Thus, these results suggest that, in general, $S_4$ is the most stable configuration in the synchronous regime, but which one becomes the final long time configuration depends on the frequency dependent probabilities of the initial and late transitions.

Figure~\ref{fig:structresp-BD-Kb0-T1-relaxdyn} also shows an example of the configurational transitions corresponding to the asynchronous regime. As already observed for the critical frequency, the initial relaxation is to $S_5$ in all cases. Only a moderate fraction further transitions to $S_4$ within a relatively short time interval, but at such field frequency, $f_H\, t_s \approx 240$, the $S_4$ configuration is not able to follow synchronously the field rotation anymore. This point is also dicussed in more detail in the overall model comparison presented in the next Section.

\subsection{Effects of hydrodynamics and overall comparison}

Finally, we analyze the effects of hydrodynamics by considering all the terms in the dynamic Equations~\ref{eq:BrownT} to \ref{eq:netT}. Hereby, we label this full filament model as `HD'. In order to ease model comparisons, we stick to the conditions chosen for the BD model: fixed very low rigidity, $l_p/L \approx 10$, and moderate field strength, $H/M_s \approx 0.25$. We also expect hydrodynamic effects to have a strong impact under these conditions.

\subsubsection{Frequency response with hydrodynamic interactions}
\begin{figure}[!h]
\centering
\includegraphics[width=0.95\columnwidth]{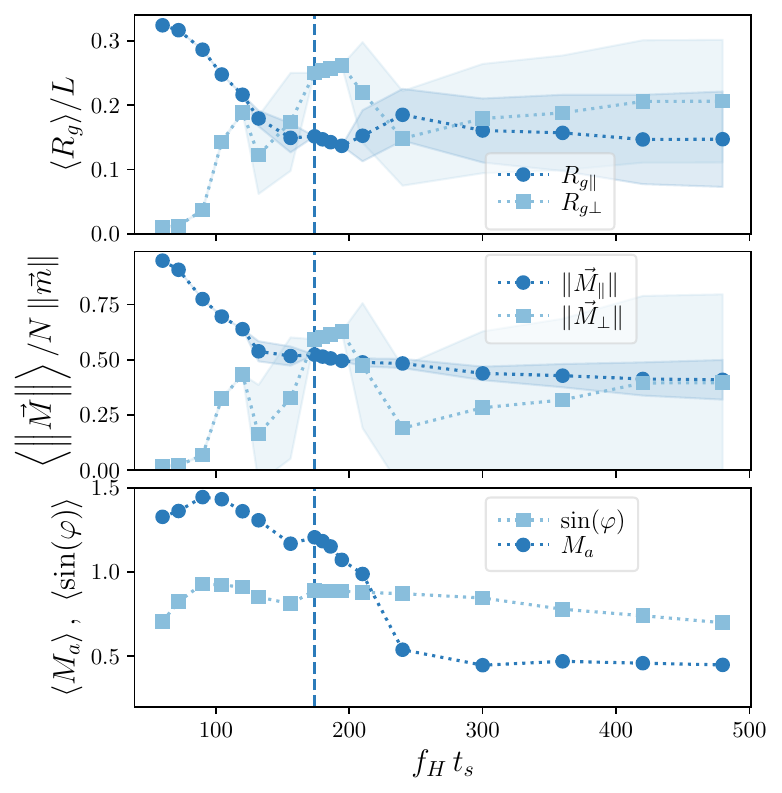}% Here is how to import EPS art
\caption{\label{fig:structresp-HD-Kb0-T1}Long-time frequency structural response---split radius of gyration, split net magnetization, phase shift and Mason number---of the HD model for a moderate field strength, $H/M_s \approx 0.25$, and low rigidity $l_p/L \approx 10$. Symbols correspond to average values, shadowed regions to their standard deviations. Vertical dashed lines show the critical frequency.}
\end{figure}
Similarly to Figures~\ref{fig:structresp-BD-T0} and \ref{fig:structresp-BD-Kb0-T1}, Figure~\ref{fig:structresp-HD-Kb0-T1} shows the dependence on the field frequency of the filament structural parameters for the HD filament model. One can see that, by introducing hydrodynamic interactions, a further shift of the critical frequency towards a higher value, $f_c \, t_s \approx 174$, is obtained. This is almost 1.4 times larger than the corresponding to the BD model, thus the impact of hydrodynamics is also quite significant. Besides a general shift of the characteristic frequencies, the parameters show similar qualitative trends as in the BD model, but here the critical frequency is not so easily signaled by their average values. For instance, the Mason number does not show a large drop at the critical frequency, as its definition does not capture the long range correlations between different parts of the filament body introduced by the hydrodynamic interactions. However, the critical frequency does correspond to the lower boundary of a narrow intermediate region in which large fluctuations of the averages vanish, indicating that only one type of configuration is found at long times. Further details are discussed in the next Section.

\subsubsection{Overall response model comparison}
\begin{figure}[!h]
\centering
\begin{subfigure}[b]{0.95\columnwidth}
\centering
\includegraphics[width=0.95\columnwidth]{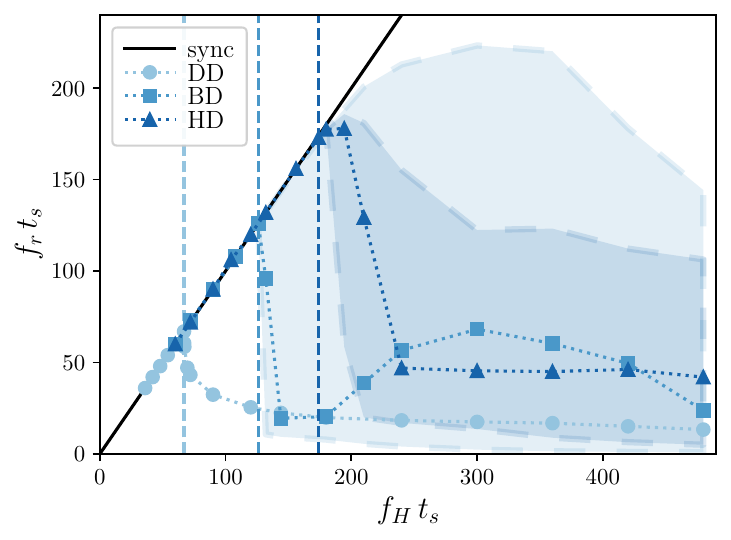}
\caption{\label{fig:meanresp-models}Filament rotation frequency as a function of field frequency for each model. Symbols correspond to average values. Dashed curves at the contours of the shadowed regions for the BD and HD models represent the frequency modes associated to the probability distributions of the selected configurations. Shadowed regions are a guide for the eye to connect symbols and contour curves. Vertical dashed lines indicate the critical frequency.}
\end{subfigure}
\begin{subfigure}[b]{0.95\columnwidth}
\centering
\includegraphics[width=0.95\columnwidth]{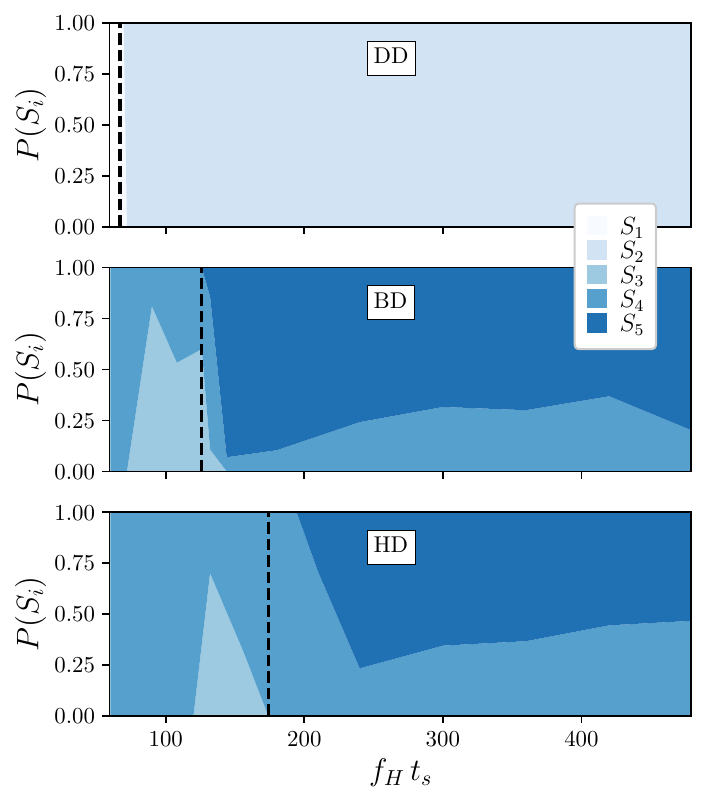}
\caption{\label{fig:splitresp-models}Stacked probability plots of each characteristic filament configuration as a function of field frequency and for each filament model. Vertical dashed lines correspond to the critical frequencies.}
\end{subfigure}
\caption{\label{fig:resp-models}Compared long-time stationary frequency responses of DD, BD and HD models for a moderate field strength, $H/M_s \approx 0.25$, and low rigidity $l_p/L \approx 10$.}
\end{figure}
To complete the analysis, we focus on the overall comparison of the filament frequency response for the three levels of modeling details we considered and the choice of parameters discussed above, $l_p / L \approx 10$ and $H/M \approx 0.25$. Figure~\ref{fig:resp-models} summarizes this comparison, showing the filament rotation frequency and the distribution probabilities of the long-time filament configurations as a function of the field frequency. The first property, shown in Figure~\ref{fig:meanresp-models}, not only displays the limit between synchronous and asynchronous response for each model but also outlines the probabilistic nature of the asynchronous behavior of BD and HD systems in the following way. In both models, such probabilistic response is strongly bimodal, with very narrow high- and low-frequency modes corresponding, respectively, to the relatively fast rotating $S_3$ or $S_4$ configurations and to the relatively slow rotating $S_5$ ones. As a simple and compact way to display such information, in Figure~\ref{fig:meanresp-models} symbols represent the averaged values of the frequency response and the curves at the contours of the shadowed regions around them indicate the high and low frequency peaks of the underlying response probability distributions. The long time probability of each particular configuration, $S_1$ to $S_5$, is shown in Figure~\ref{fig:splitresp-models} by means of stacked probability plots.

Figures~\ref{fig:meanresp-models} and \ref{fig:splitresp-models} clarify the strong and complex impact, outlined in Sections above, of thermal fluctuations and hydrodynamic interactions on the filament response. Whereas the main differences between the results for the deterministic and the two thermalized models have been already established, differences between the latter still need a careful discussion. In the synchronous regime we can observe not only a shift of the critical frequency but also some difference in the way fractions of $S_3$ and $S_4$ configurations depend on the field frequency (see Figure~\ref{fig:splitresp-models}). For the BD model, the onset of $S_3$ takes place at lower $f_H$ and grows to a larger fraction than for the HD case. In both models it reaches a maximum and decays as the critical frequency is approached. For the BD system under its $f_c$, however, $S_3$ fraction is still slightly more likely than $S_4$ one, whereas the critical frequency of the HD model is above the synchronous limit of $S_3$ and only $S_4$ configurations are found. All this suggests that long-ranged hydrodynamic interactions not only help synchronicity but also tend to favor more overall extended configurations. In the asynchronous regime, the BD model shows a very abrupt drop of the average frequency response right above the critical value (see Figure~\ref{fig:meanresp-models}), corresponding to the onset of the $S_5$ configurations, whose fraction rapidly becomes dominant as $f_H$ increases (see Figure~\ref{fig:splitresp-models}). However, we can see that for any sampled frequency there is a nonzero $S_4$ fraction, as well as a small $S_3$ fraction within a narrow range right above the critical frequency. These configurations are still able to perform synchronous rotations within a significant interval of the strictly defined asynchronous region. This is indicated in Figure~\ref{fig:meanresp-models} by the upper contour curve of the BD shadowed region laying on the ideal synchronous line. Interestingly, the maximum frequency at which synchronous rotations are statistically possible in the BD model corresponds approximately to the critical frequency of the HD model. For the latter we can see that the onset of $S_5$ configurations takes place at a frequency well above the critical limit and the growth of its fraction with frequency is less abrupt than in the BD case. In addition, for the HD model there is a zero probability of obtaining a synchronous response at any frequency above its critical limit. Therefore, the critical frequency of the HD model represents the upper limit for $S_4$ configurations to be able to perform synchronous rotations in both models. For frequencies higher than that, BD and HD systems also behave differently. $S_4$ configurations in the BD model are still able to increase the frequency of their asynchronous response up to an absolute maximum around $f_H\, t_s \approx 300$, to finally drop significantly at higher field frequencies. On the contrary, $S_4$ configurations in the HD model tend to monotonically decrease their response values in practically the entire range of field frequencies above the critical limit. Another difference between both models lies in the response values of $S_5$ configurations, represented by the lower contour curves of the shadowed regions in Figure~\ref{fig:meanresp-models}: while for the BD model there is a drop of the response to almost zero right above the critical frequency, in the HD model the corresponding drop is slightly less pronounced. Thus, to this respect, the general effect is that hydrodynamic interactions tend to narrow the bimodal distribution of aynchronous response values introduced by thermal fluctuations.

\section{Conclusions and outlook}
We characterized the response of a magnetic filament, made of ferromagnetic nanoparticles in the limit of infinite magnetic anisotropy, when exposed to rotating magnetic fields. By means of nonequilibrium molecular dynamics simulations, we explored for the first time the behavior of such system at the nanoscale, as well as in the limit of high filament backbone flexibilities.

In order to determine the qualitative impact of filament flexibility, thermal fluctuations and hydrodynamic interactions on the frequency response, we performed different sets of simulations with an increasing level of modeling details. For an athermal system without hydrodynamic interactions, we recover the known behavior of microparticle filament models in the limit of large rigidities, whereas in the limit of high flexibility we observe a broadening of the synchronous frequency response regime and the appearance of large twodimensional structural oscillations in the asynchronous regime. When thermal fluctuations are considered for the case of a highly flexible filament, its structural behavior becomes probabilistic, being able to adopt three characteristic threedimensional configurations with two associated rotation frequencies. This takes place by means of an initial fast relaxation and a later eventual transition to the final dynamic configuration, following frequency dependent probability distributions for such processes. As a result, there is a further broadening of the synchrnous regime and a strong and broadly separated bimodal probability distribution for the asynchronous frequency response. Finally, hydrodynamic interactions perturb significantly the latter behavior by, first, shifting still further the limit of the synchronous regime towards higher frequencies and, second, narrowing the difference between the two characteristic frequencies of the bimodal distribution observed in the asynchronous regime.

Despite the apparent simplicity of this system, our results indicate its rather complex nonequilibrium behavior, which we hope will stimulate future works. For instance, natural extensions of this work are the study of the effects of finite magnetic anisotropies for the filament particles and a more accurate treatment of hydrodynamic interactions that could pave the way for quantitative studies of more complex systems, such as nanofilament dispersions or confined systems. Also particularly interesting would be the study of the behavior of not fully rigid MFs in viscoelastic environments, including biological systems.

\begin{acknowledgments}
PAS acknowledges support from the POSTDOC-UIB-2020 project "Computer modeling of magnetic nanosorbents", funded by the University of the Balearic Islands and the European Regional Development Fund.
\end{acknowledgments}

\end{document}